\documentclass[superscriptaddress,twocolumn,secnumarabic,
amssymb,amsmath,nobibnotes,aps,prd,nofootinbib]{aastex62}
\graphicspath{{./}{figures/}}
\usepackage[utf8]{inputenc}
\usepackage{graphicx}% Include figure file
\usepackage{dcolumn}% Align table columns on decimal point
\usepackage{bm}% bold math
\usepackage{ulem} %here for \sout, removed to avoid funny underlines
\usepackage{hyperref}% add hypertext capabilities
\usepackage{multirow,enumitem}
\usepackage{amsmath}
\usepackage{color}
\usepackage{xcolor}
\usepackage{multirow}

\newcommand{\Om}{\Omega_m}
\newcommand{\LCDM}{\rm{\Lambda}CDM}

\begin{document}
\title{HUBBLE DIAGRAM AT HIGHER REDSHIFTS: MODEL INDEPENDENT CALIBRATION OF QUASARS }

\author{Xiaolei Li}
\altaffiliation{lixiaolei@hebtu.edu.cn}
\affiliation{College of Physics, Hebei Normal University, Shijiazhuang 050024, China}

\author{Ryan E. Keeley}
\altaffiliation{rkeeley@kasi.re.kr}
\affiliation{Korea Astronomy and Space Science Institute, Daejeon 34055, Korea}
\author{Arman Shafieloo}
\altaffiliation{shafieloo@kasi.re.kr}
\affiliation{Korea Astronomy and Space Science Institute, Daejeon 34055, Korea}
\affiliation{University of Science and Technology, Yuseong-gu 217 Gajeong-ro, Daejeon 34113, Korea}
\author{Xiaogang Zheng}
\affiliation{School of Electrical and Electronic Engineering,
Wuhan Polytechnic University, Wuhan 430023, China}
\author{Shuo Cao}
\affiliation{Department of Astronomy, Beijing Normal University, Beijing 100875, China}
\author{Marek Biesiada}
\affiliation{Department of Astronomy, Beijing Normal University, Beijing 100875, China}
\affiliation{National Centre for Nuclear Research, Pasteura 7, 02-093
Warsaw, Poland}
\author{Zong-Hong Zhu}
\affiliation{Department of Astronomy, Beijing Normal University, Beijing 100875, China}
\date{\today}% It is always \today, today,
             %  but any date may be explicitly specified
 
\begin{abstract}
In this paper, we present a model-independent approach 
to calibrate the largest quasar sample. Calibrating quasar samples is essentially constraining the parameters of the linear relation between the $\log$ of the ultraviolet (UV) and X-ray luminosities. This calibration allows quasars to be used as standardized candles. There is a strong correlation between the parameters characterizing the quasar luminosity relation and the cosmological distances inferred from using quasars as standardized candles. We break this degeneracy by using Gaussian process regression to model-independently reconstruct the expansion history of the Universe from the latest type Ia supernova observations. Using the calibrated quasar dataset, we further reconstruct the expansion history up to redshift of $z\sim 7.5$. Finally, we test the consistency between the calibrated quasar sample and the standard $\LCDM$ model based on the posterior probability distribution of the GP hyperparameters. Our results show that the quasar sample is in good agreement with the standard $\LCDM$ model in the redshift range of the supernova, despite of mildly significant deviations taking place at higher redshifts. Fitting the standard $\LCDM$ model to the calibrated quasar sample, we obtain a high value of the matter density parameter $\Omega_m = 0.382^{+0.045}_{-0.042}$, which is marginally consistent with the constraints from other cosmological observations. 
 
\end{abstract}

\section{Introduction} \label{sec:intro}

As a potential cosmic probe at higher redshifts, quasars might be able to fill the redshift gap between the farthest observed Type Ia supernovae (SN Ia) \citep{scolnic2017complete} and the cosmic microwave background (CMB) \citep{Aghanim:2018eyx} owing to the fact that quasars are luminous persistent sources in the Universe and can be observed up to redshifts of $z\approx7.5$ \citep{Mortlock:2011va}. For instance, recently it has been proposed that intermediate-luminosity radio quasars could potentially provide a new type of standard rulers, which extended our understanding of the evolution of the Universe to $z\sim 3$ \citep{2017JCAP0012C,2017A&A...606A..15C,2018EPJC...78..749C,2019MNRAS.483.1104Q,Qi_2021,2017arXiv170808867L}. More interestingly, quasars have also been used as standard candles whose standardization relies on the linear relation between the $\log$ of their ultraviolet (UV) and X-ray luminosities \citep{Risaliti:2015zla,Risaliti:2016nqt,lusso2016tight,Lusso:2017hgz,Risaliti:2018reu,Salvestrini:2019thn,Lusso:2019akb,Lusso:2020pdb,Lusso:2020obu,Khadka:2020vlh,Liu:2020pfa,Liutonghua_20201,2020MNRAS.496..708L,2020ApJ...905...54G,Zheng:2021}. 

So far, the largest quasar sample with both X-ray and UV observations consists of $\sim 12,000$ objects, assembled by combining several different samples in \citet{Lusso:2020pdb}. The full sample includes 29 quasars from XMM-Newton at $z\simeq 3$ \citep {Nardini:2019}, 1 new optically-selected quasar at $z\sim 4$ from XMM-Newton \citep{Nardini:2019}, 64 high-$z$ quasars from \citet{Salvestrini:2019thn}, 840 quasars from the XMM-XLL sample published by \citet{10.1093/mnras/stv2749}, 9252 quasars from the SDSS-4XMM sample \citep{Paris:2016xdm,Mingo:2016pdq,Webb:2020rgy}, 2392 quasars from \citet{Paris:2016xdm,Evans:2010ye}, and 15 local AGN selected in \citet{Lusso:2020pdb}. Note that several filtering steps were applied to reduce the systematic effects and 2421 quasars in the redshift range $0.009<z<7.5$ were left in the final cleaned sample \citep{Lusso:2020pdb}. The relation between the X-ray and UV luminosities is usually parameterized as $ {\log}(L_{\rm{X}})\,=\,\gamma {\log}(L_{\rm{UV}})+\beta_1$, where $L_{\rm{X}}$ and $L_{\rm{UV}}$ are the rest-frame monochromatic luminosities at 2 keV and 2500 \AA, respectively. There is a total of 3 free parameters for the quasar calibration, the slope of the relation $\gamma$, the offset of the relation $\beta_1$, and the intrinsic dispersion $\delta$, which is not 
shown directly in the equation.

Up to now, different methods have been used to calibrate quasar samples. In \cite{Risaliti:2015zla}, the authors obtained the best-fit values of $\gamma$ and $\beta_1$ based on the $\LCDM$ model. The parameter $\beta_1$ shifts the relation between the UV and X-ray luminosities up and down and thus also scales the corresponding distance-redshift relation up and down. When doing cosmological inference, $\beta_1$ is thus degenerate with Hubble constant $H_0$ and only a combination of the two can be measured but not either individually. One way to calculate the value of $\gamma$ was carried out in \cite{Risaliti:2018reu}, in which the authors split the quasar sample into subsamples within narrow redshift bins and then fit the linear $\log (F_{\rm{UV}})- \log (F_{\rm{X}})$ relation in each redshift bin. In this way, an average value of $\gamma$ was obtained in the end. In this work, we introduce a novel model-independent technique to calibrate the quasar sample, in which we use Gaussian process (GP) regression to reconstruct the expansion history from the Pantheon SN Ia dataset.  We then use these reconstructions of the unanchored luminosity distances ($D_{\rm{L}}H_0$) to break the degeneracy between the parameters of the quasar calibration and the expansion history, thus calibrating the quasars in a way that are, by construction, consistent with the SN Ia and independent of any cosmological or parametric assumption. It should be emphasised here that, all the work is done based on the assumption that there is no evolution of the $L_{\rm{UV}}-L_{\rm{X}}$ relation with redshift and this assumption has been tested in the previous works.

This paper is organised as follows, in Section~\ref{sec:cali}, we describe the model-independent quasar calibration method in detail and show the Hubble diagram of the calibrated quasars. We test the reliability of the calibration results in Section~\ref{sec:cross_ckeck}. In Section~\ref{sec:cosmology}, we constrain the $\LCDM$ model with the calibrated quasar sample. We reconstruct the expansion history with GP from the calibrated quasar sample with the best-fit $\LCDM$ model as a mean function and test the consistency between the calibrated quasar sample and the $\LCDM$ model in Section~\ref{sec:consistency}. We discuss our conclusions in Section~\ref{sec:conclusion}.

\section{Calibration} \label{sec:cali}

\begin{figure*}[ht!]
\centering
\includegraphics[width=0.875\textwidth]{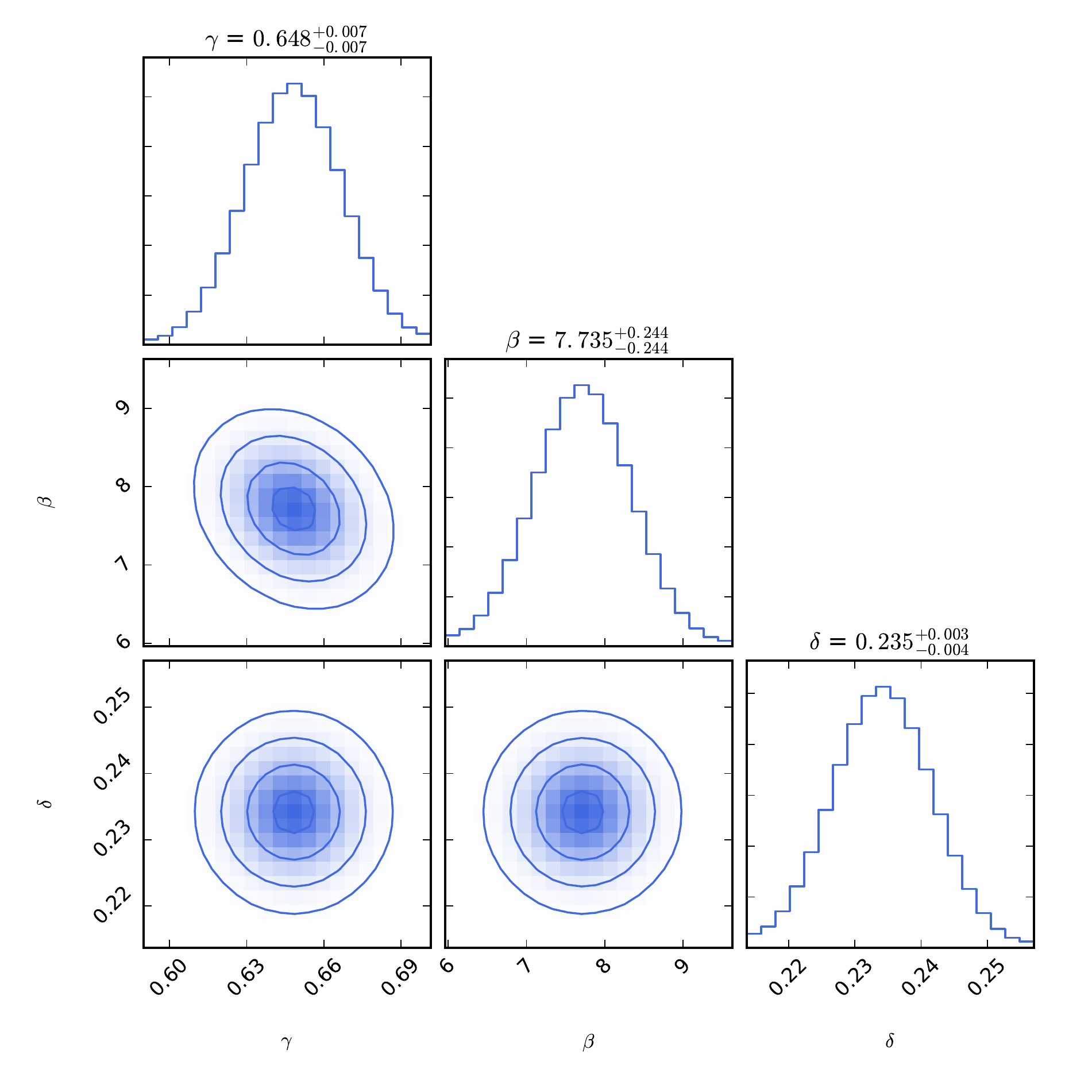}
\caption{Model-independent calibration results for the quasar parameters. GP reconstructions of $D_LH_0$ based on the Pantheon SN Ia compilation were used.
The contours represent the 1$\sigma$, 2$\sigma$, and 3$\sigma$ uncertainties for $\gamma , \beta,$ and $\delta$.
Marginal distributions for each parameter are shown on the top of each 2-D subplot.}
\label{fig:quasar_calibration_res}
\end{figure*}

Here, we calibrate the 2421 quasar sample, which was compiled in \cite{Lusso:2020pdb}.
In this section, we describe the calibration method in detail and show the calibration results. Moreover, the reliability of the calibration results are also discussed in the next section.

From the linear relation between $\log(L_{{\rm{UV}}})$ and $\log(L_{\rm{X}})$, $ {\rm{log }}(L_{\rm{X}})\,=\,\gamma {\rm{log }}(L_{\rm{UV}})+\beta_1$, one can obtain
\begin{equation} \label{eq:logFx1}
    {\log (F_{\rm{X}})}\,=\,\gamma \log (F_{\rm{UV}}) + (2\gamma-2){\rm{log }}(D_{\rm{L}}) + \beta_2
\end{equation}
where $\beta_2 = \gamma \log (4\pi)-\log (4\pi)+\beta_1$, $F_{\rm{UV}}$ and $F_{\rm{X}}$ are the fluxes measured at fixed rest-frame wavelengths, and $D_{\rm{L}}$ is the luminosity distance. In order to calibrate the quasar parameters in a model-independent way, an external observation is needed. In other words, since the quasar calibration parameters are degenerate with the cosmological distances, if we use the cosmological distances from another tracer of the expansion history (i.e. SN Ia), we can break this degeneracy and tightly constrain the calibration parameters. During our work, we generate samples of the unanchored luminosity distance $D_{\rm{L}}H_0$ from the posterior of the Pantheon compilation from \cite{scolnic2017complete} calculated with GP (see \cite{KeeleySLa,KeeleySLb} for details on this sampling and see \cite{Rasmussen:2006,Holsclaw:2010nb,Holsclaw:2010sk,2011PhRvD..84h3501H,Shafieloo2012Gaussian,KeeleyGP,KeeleyTDE,KeeleyGW,KeeleyeBOSS} for a broader discussion of GP).

As a reminder, since the absolute brightness of the SN Ia is degenerate with $H_0$, only the dimensionless, unanchored luminosity distances ($D_LH_0$) can be measured.

GP regression works by generating a set of cosmological functions from the covariance function between the values at different redshifts. We follow some previous works and assume that the covariance function is parameterized as a squared-exponential kernel \citep{Rasmussen:2006,Holsclaw:2010nb,Holsclaw:2010sk,2011PhRvD..84h3501H,Shafieloo2012Gaussian} 
\begin{equation}\label{eq:kernel}
   <\varphi(s_i)\varphi(s_j)>\,=\,\sigma_f^2 \exp\left({-\frac{|s_i-s_j|^2}{2\ell^2}}\right)
\end{equation}
where $s_i=\ln (1+z_i)/\ln (1+z_{\rm{max}})$ and $z_{\rm{max}}=2.26$ is the maximum redshift of the SN Ia sample. There are two hyperparameters, $\sigma_f$ and $\ell$, that are marginalized over. These hyperparameters determine the amplitude of the random fluctuations and the coherence length of the fluctuation, respectively. 
$\varphi$ is just a random function drawn from the distribution defined by the covariance function of equation.~(\ref{eq:kernel}) and we take this function as $\varphi(z)=\ln \left(H^{\rm{mf}}(z)/H(z)  \right)$, i.e. the logarithm of the ratio  between the reconstructed expansion history, $H(z)$, and a mean function, $H^{\rm{mf}}(z)$, which we choose to be the best-fit $\LCDM$ model from Pantheon dataset.
The mean function plays an important role in GP regression and the final reconstruction results are not quite independent of the mean function, however it has a modest effect on the final reconstruction results because the values of hyperparameters help to trace the deviations from the mean function \citep{Shafieloo2012Gaussian,2013PhRvD..87b3520S,2017JCAP...09..031A}. 
Moreover, the true model should be very close to the flat $\LCDM$ model so it is reasonable to choose the best-fit flat $\LCDM$ model from Pantheon as a mean function.

With the reconstructed expansion history $H(z)$, 
we can integrate this function to get the unanchored luminosity distance,
\begin{equation}
    D_LH_0(z) = (1+z)\int_0^z dz \frac{c}{h(z)}
\end{equation}
where $h(z) = H(z)/H_0$.  It is this function that is most directly constrained by the SN Ia data and can thus be reconstructed. GP effectively calculates a posterior for this function, 
\begin{equation}
    P(D_LH_0(z)|D) = \int d\varphi \mathcal{L}(D_LH_0(\varphi)) P(\varphi)/P(D),
\end{equation} 
using the likelihood of the data $\mathcal{L}(D_LH_0(\varphi))$ (see equation~(\ref{eq:chi2})) and a prior on $\varphi$ (which is a consequence using a flat prior on the GP hyperparameters $\sigma_f$ and $\ell$).
It is from this posterior distribution that we draw samples of the unanchored luminosity distance.

With the GP results in hand, the first step in calibrating the quasar distances is to draw 1000 unanchored luminosity distances $D_{\rm{L}}H_0$ reconstructed from the SN Ia data. We then calculate the predicted quasar X-ray flux corresponding to these unanchored luminosity distances $D_{\rm{L}}H_0$ by rewriting equation~(\ref{eq:logFx1}) as,
\begin{equation} \label{eq:logFx}
    \log (F_{\rm{X}})\,=\, \gamma \log (F_{\rm{UV}}) + (2\gamma-2){\rm{log }}(D_{\rm{L}}H_0) + \beta
\end{equation}
where $\beta=\beta_2-(2\gamma-2)\log (H_0)$. 
With the measurements of $F_{\rm{UV}}$ and the $D_{\rm{L}}H_0$ from SN Ia, we obtain ${\rm{log }}(F_{\rm{X}})^{\rm{SN}}$ following equation~(\ref{eq:logFx}). This allows us to compare the quasar dataset and the SN Ia dataset.

Then, following \cite{Risaliti:2015zla,Lusso:2020pdb}, we define the likelihood ($\mathcal{L} = \exp\left( -\chi^2 /2 \right)$) of the quasar parameters based on a modified $\chi^2$ function, which includes a penalty term for the intrinsic dispersion $\delta$ 
\begin{equation}\label{eq:chi2}
\begin{aligned}
    \chi^2\,=\,\sum_i [& \frac{\left(\log (F_{\rm{X}}(\gamma,\beta))^{\rm{SN}}_i- \log (F_{\rm{X}})_i^{\rm{QSO}}\right)^2} {s_i^2} \\
   &  +{\rm{ln}} (s_i^2)]
\end{aligned}
\end{equation}
where $s_i^2\,=\,\sigma_{\log (F_{\rm{X}})}^2+\gamma^2 \sigma_{\log (F_{\rm{UV}})}^2+\delta^2$. 
The intrinsic dispersion $\delta$ of the $L_{\rm{L}}-L_{\rm{UV}}$ relation is considered in order to reduce the Eddington bias which has the effect of flattening the $L_{\rm{L}}-L_{\rm{UV}}$ relation \citep{Risaliti:2018reu,Lusso:2020pdb}. 
Further, a non-zero value for this parameter is needed in order yield a reasonable $\chi^2$ per degree of freedom.  The observed dispersion of the the X-ray fluxes is far larger than just the measurement error, so there must be some additional variance that is an intrinsic feature of the quasar population, and not the measurement of them. We account for this intrinsic scatter with the parameter $\delta$.
With equation~(\ref{eq:chi2}), we use the $\bf{LINMIX\_ERR}$ method \citep{Kelly:2007jy}, which 
accounts for measurements uncertainties on both independent and dependent variables, non-detections, and intrinsic scatter. The penalty term is important to guard against the intrinsic dispersion growing too large in the fit. For example, simply minimizing the $\chi^2$ can be trivially achieved with a sufficiently large value of $\delta$. However, the $\chi^2$ per degree of freedom in this case would be close to zero, not one. Thus the penalty term assures that $\delta$ is only as large as it needs to be to make the residuals Gaussian distributed ($\chi^2$ per degree of freedom $\sim 1$).  
We then calculate the posterior distribution of the quasar parameters: the slope $\gamma$, the intercept $\beta$ and the intrinsic dispersion parameter $\delta$. We should note that the Hubble constant $H_0$ is absorbed into the parameter $\beta$. 
Based on the method described above, we use a Python package named \textit{emcee} \citep{foreman2013emcee} to do the MCMC analysis and flat priors are used for each parameter.

Our calibration method can be summarized as follows:
\begin{enumerate}
  \item Draw 1000 unanchored luminosity distances $D_{\rm{L}}H_0$ from supernovae data,
  \item Calculate the predicted quasar X-ray flux corresponding to these unanchored luminosity distances,
\item Define the likelihood of the quasar parameters,
\item Calculate the posterior distribution of the quasar parameters.
\end{enumerate}

The best fit values for quasar parameters $\gamma$, $\beta$, and $\delta$ and 1$\sigma$, 2$\sigma$, and 3$\sigma$ uncertainties are shown in Figure~\ref{fig:quasar_calibration_res}. We find that $\gamma = 0.649 \pm 0.007$, and $\delta = 0.235 \pm 0.04$.
One can see that our constraints on the slope parameter $\gamma$ and the intrinsic dispersion parameter are consistent at the 1$\sigma$ confidence level with the calibration results from \cite{Lusso:2020pdb} which gives $\gamma\,=\,0.586\pm0.061$, $\delta\,=\,0.21\pm0.06$ by dividing the sample in redshift bins and fitting the $F_{\rm{X}}-F_{\rm{UV}}$ relation in the chosen redshift bins.
The intercept parameter $\beta$ is a function of both $\beta_1$ and $H_0$ and represents the relative anchoring between the unanchored luminosity distances and the observed quasar fluxes.  It has absorbed all of the information about the anchoring of the data and other anchoring of the model.

\section{Testing the internal consistencies} \label{sec:cross_ckeck}

\begin{figure}[t]
\centering
\includegraphics[width=\columnwidth]{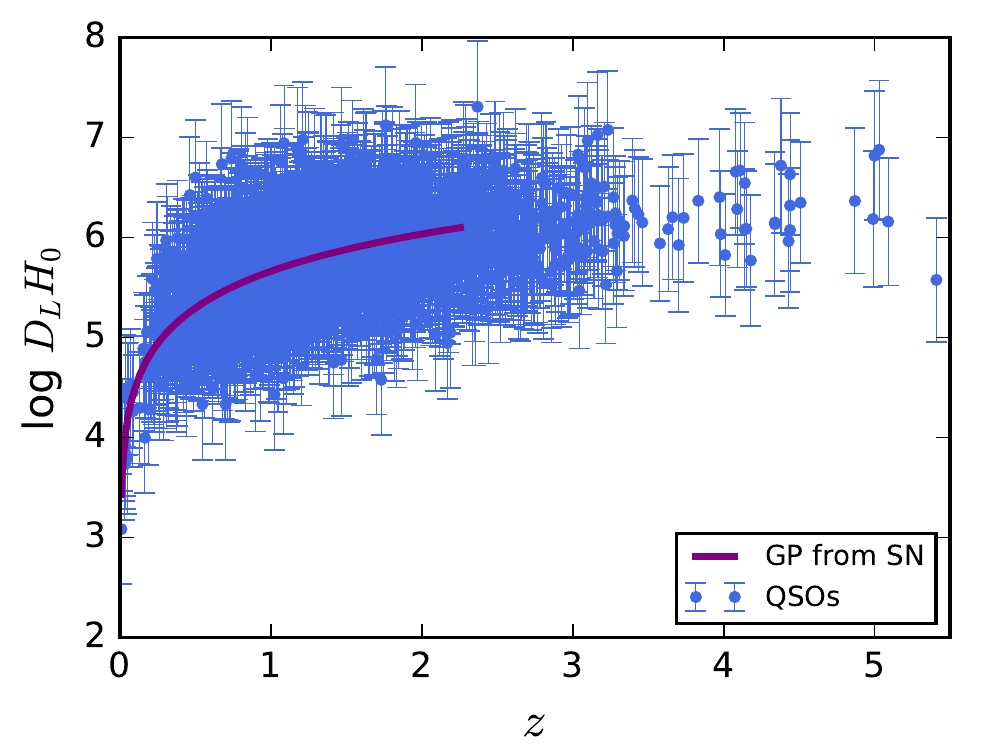} 
\caption{$\log (D_{\rm{L}}H_{\rm{0}})$-redshift relation for the 2421 calibrated quasars. The errorbars of $\log (D_{\rm{L}}H_{\rm{0}})$ are obtained through error propagation and the purple solid line shows $\log (D_{\rm{L}}H_0)$ drawn from the posterior of the Pantheon compilation calculated with GP.}
\label{fig:DLH0_from_QSOs}
\end{figure}

With the best fits for the quasar parameters $\gamma$, $\beta$ and $\delta$, we can then calculate a series of checks to be sure the calibration results infer reasonable information about cosmology. For instance, the $\log (D_{\rm{L}}H_0)$ vs. $z$ relation can be obtained from the quasar fluxes and calibrated quasar parameters via 
\begin{equation} \label{eq:logDLH0}
{\rm{log}}(D_{\rm{L}}H_0)\,=\,\frac{\log (F_{\rm{X}})-\gamma \log (F_{\rm{UV}})-\beta}{(2\gamma-2)}.
\end{equation} 
Figure~\ref{fig:DLH0_from_QSOs} shows the $\log (D_{\rm{L}}H_0)$ vs. $z$ relation for the 2421 quasar sample, together with the median inference from the posterior of the Pantheon compilation calculated with GP regression, which is shown by the purple solid line. The blue points represent the quasar data transformed from fluxes to unanchored luminosity distances via equation~(\ref{eq:logDLH0}) along with the uncertainties obtained according to error propagation. These calibrated, unanchored quasar luminosity distances comprises the dataset that we use to make inferences about cosmology
in later sections.

\begin{figure}[t]
\centering
\includegraphics[width=0.49\textwidth]{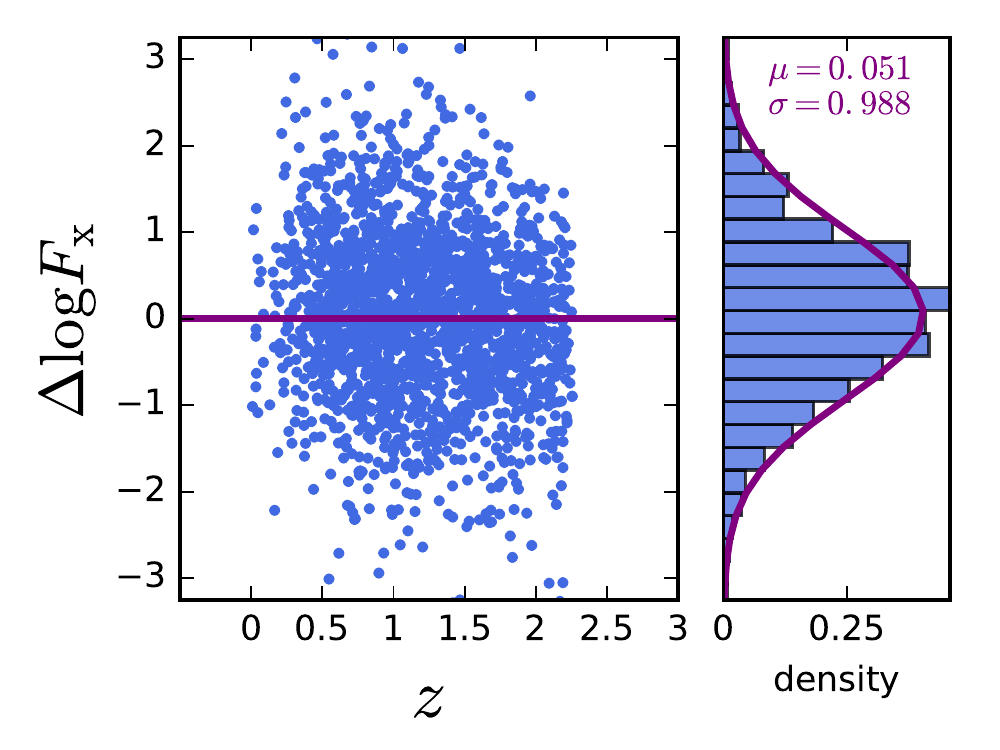}  
\caption{Residuals of the observed $\log  (F_{\rm{X}})$ values with respect to the predicted $\log  (F_{\rm{X}})$ values derived from the GP reconstructions of the Pantheon SN Ia compilation, normalized to the calibrated errors (observational and intrinsic). The right plot shows the histogram for $\Delta \log  (F_{\rm{X}})$ and the purple line shows the best Gaussian fit with $\mu = -0.051$ and $\sigma=0.988$. }
\label{fig:residual}
\end{figure}

On the other hand, in order to test the consistency between the calibrated quasar sample and the unanchored luminosity distance from SN Ia, we adopt the best-fit values of the three quasar parameters to estimate the normalized residual of $\log (F_{\rm{X}})^{\rm{SN}}$ from equation~(\ref{eq:logFx}) with respect to the measurement of $\log  (F_{\rm{X}})$ following 
\begin{equation}\label{eq:residual}
    \Delta \log (F_{\rm{X}})\,=\,\frac{\log (F_{\rm{X}})^{\rm{SN}}-\log (F_{\rm{X}})^{\rm{{QSO}}}}{\sqrt{\sigma_{\log (F_{\rm{X}})}^2+\gamma^2 \sigma_{\log (F_{\rm{UV}})}^2+\delta^2}}.
\end{equation}
The result for the residual is shown in Figure~\ref{fig:residual}. As can be seen from the right plot of Figure~\ref{fig:residual}, the distribution of the normalized residual is a Gaussian distribution, which indicates that $\rm{log }(F_{X})$ data derived from the best-fit values of the quasar parameters in our approach is consistent with $\rm{log }(F_{X})$derived from supernovae data. Finding no evidence against the linear $\log(F_{\rm{UV}})-\log(F_{\rm{X}})$ relation the calibration method described above 
(based on Type Ia supernovae that we consider to be standardized candles) 
results in a well-behaved dataset showing internal consistency (the residuals are Gaussian distributed). We 
then use this calibrated dataset to reconstruct the expansion history.

\begin{figure}
\centering
\includegraphics[width=0.475\textwidth]{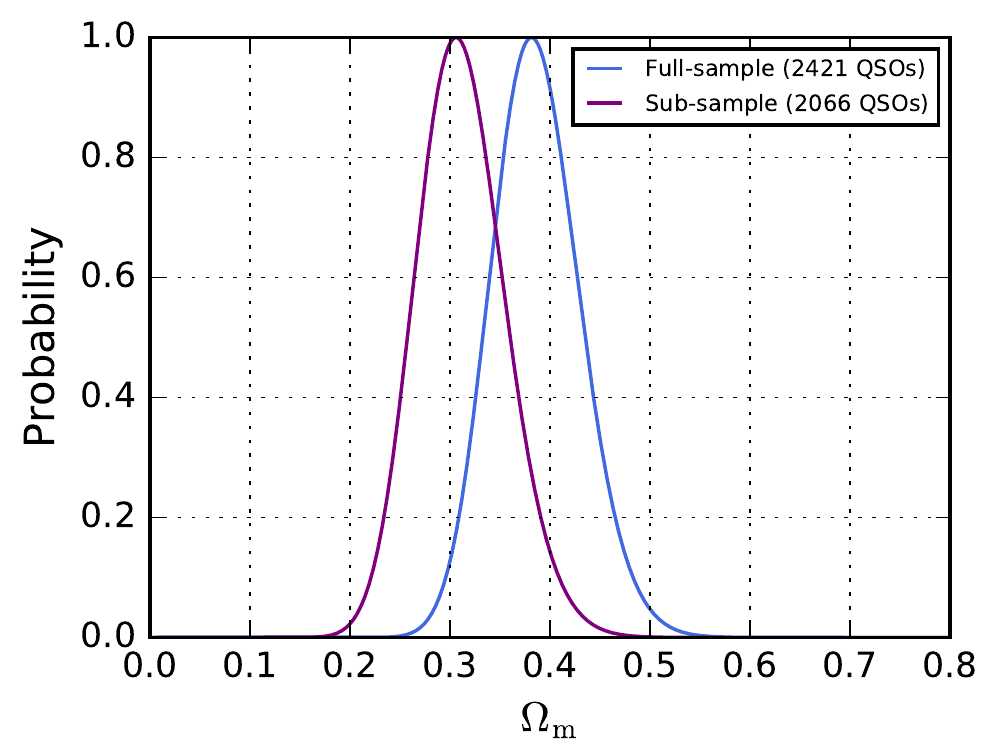}   
\caption{Constraints for a flat $\LCDM$ model from the calibrated quasar sample using both the full sample (blue) and the subsample (purple) of 2066 quasars with redshifts up to the maximum SN Ia redshift ($z\le 2.26$).}
\label{fig:best_fit_om}
\end{figure}

\section{$\Lambda$CDM Inferences}\label{sec:cosmology}
With the best-fit quasar parameters, we have transformed the quasar fluxes into a set of unanchored luminosity distances, which we can now use to constrain the $\Lambda$CDM model (or any other cosmological model).
In the flat-$\Lambda$CDM model, the unanchored luminosity distance can be written as,
\begin{equation}
    D_{\rm{L}}H_0\,=\,c(1+z) \int_0^z\frac{dz}{\sqrt{\Omega_{\rm{m}}(1+z)^3+(1-\Omega_{\rm{m}})}},
\end{equation}\label{eq:DLH0_lcdm}
where $\Omega_{\rm{m}}$ is the current matter density. With this equation we can calculate $\log (D_{\rm{L}}H_0)^{\LCDM}$. The likelihood for the calibrated quasar distance dataset $\left(\log ( D_{\rm{L}}H_0)_i^{\rm{QSO}}\right)$ is then defined as
\begin{equation}\label{eq:chi2_LCDM}
\ln{\mathcal{L}}\,=\,-\frac{1}{2}\sum_i \left[ \frac{\left(\log ( D_{\rm{L}}H_0(\Omega_m))_i^{\LCDM}- \log ( D_{\rm{L}}H_0)_i^{\rm{QSO}}\right)^2} {\sigma_{\log (D_{\rm{L}}H_0);i}^2} \right].
  \end{equation}
With this likelihood function we can then calculate the posterior for $\Omega_{\rm m}$ for both the entire calibrated quasar dataset, and the subsample of quasars which consists of 2066 quasars with redshifts up to the largest redshift of SN Ia ($z=2.3$).
The constraints on $\Om$ are shown in Figure~\ref{fig:best_fit_om}. With the full quasar sample, we get $\Omega_m = 0.382^{+0.045}_{-0.042}$ and with the subsample we get $\Omega_m = 0.306^{+0.046}_{-0.042}$. The best-fit result of $\Om$ from the subsample of quasars is highly consistent with the results from \citet{Risaliti:2018reu} ($\Omega_m=0.31\pm0.05$). However, the fits on the matter density parameter from the full quasar sample show significant deviations from the value of $\Om\,=\,0.3$. The fact that we could recover $\Omega_{\rm} = 0.306$ from the subsample of the calibrated quasar dataset is relatively trivial, considering the fact that the quasars are calibrated with the Pantheon compilation which also infers $\Omega_{\rm m} \sim 0.3$ in the framework of flat-$\Lambda$CDM model. Hence it makes no surprise that the calibrated quasar dataset returns the same value when using only the quasars in the range of the Pantheon dataset. What is more interesting is the quasars at higher redshift will shift the best-fit $\Omega_{\rm m}$ to higher values, when the full quasar sample is taken into account.

\begin{figure}[t]
\centering
\includegraphics[width=0.49\textwidth]{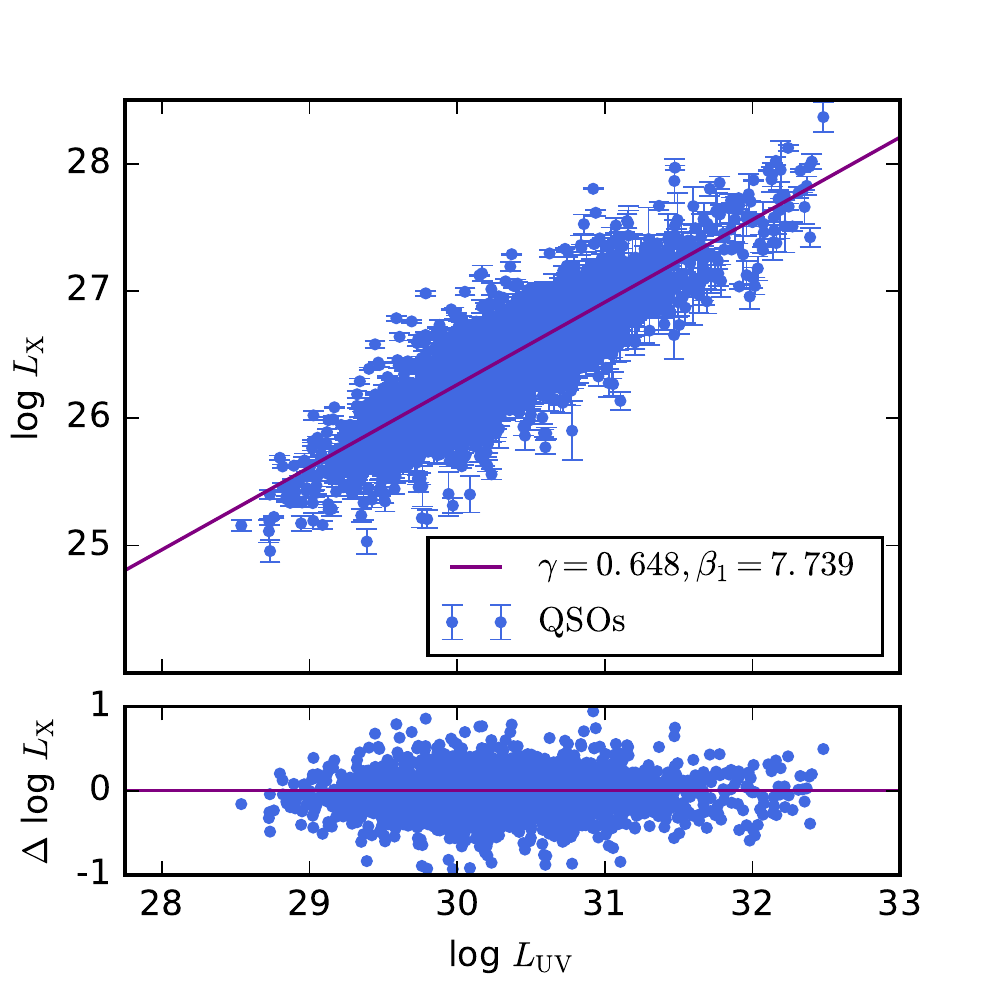}   
\caption{The linear relation between $\log({L_{\rm{UV}}})$ and $\log({L_{\rm{X}}})$ for the 2421 quasar sample we used. The purple solid line presents the best fit from our calibration results with slope $\gamma = 0.648$. The lower panel shows the residual, $\Delta \log (L_{\rm{X}})$ with respect to the best fit results.}
\label{fig:LuvLx}
\end{figure}

Based on the best-fits results from flat $\LCDM$ model, we can calculate the luminosity of the quasars from their flux measurements and the luminosity distances from the $\Lambda$CDM model with $\Omega_m = 0.382$, $\Omega_{\rm{K}}=0$ for the full sample of 2421 quasars. The results are shown in Figure~\ref{fig:LuvLx} and the purple solid line is calculated from the best-fit quasar parameters from our calibration results. The lower panel shows the residuals $\Delta \log (L_{\rm{X}})$, with respect to the best fit quasar parameters. Figure~\ref{fig:LuvLx} demonstrate the linear relation between $\log({L_{\rm{UV}}})$ and $\log({L_{\rm{X}}})$ for quasars based on our calibration results and assumption of the standard $\LCDM$ model. While visually everything seems to be consistent withe each other, more quantitative inspections are needed to check the consistency of the $\LCDM$ model and the data at the higher redshifts (redshifts beyond the range of the calibration).

\section{GP reconstruction and consistency test} \label{sec:consistency}

\begin{figure*}[t]
\centering
\includegraphics[width=\columnwidth]{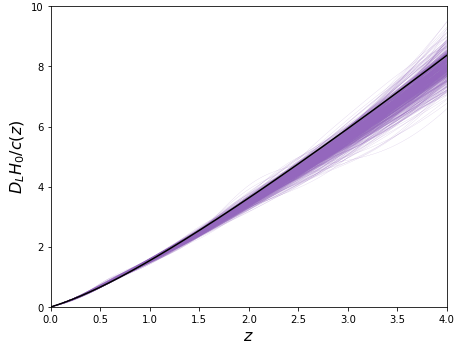}
\includegraphics[width=\columnwidth]{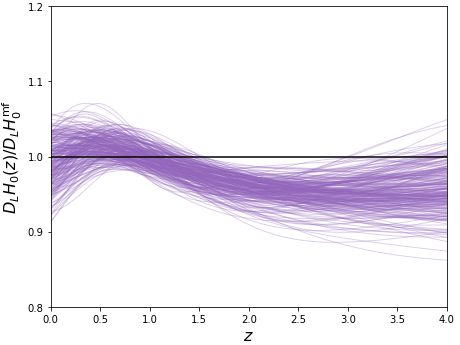}
\includegraphics[width=\columnwidth]{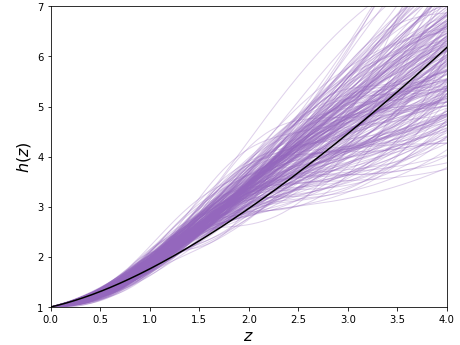}
\includegraphics[width=\columnwidth]{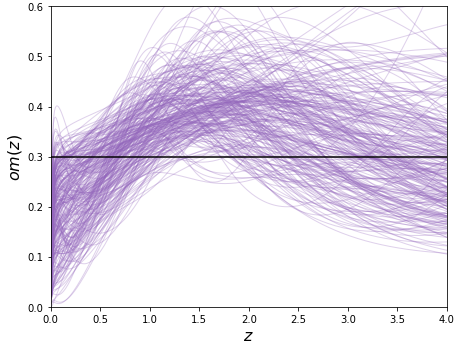}
\caption{GP reconstructions with $\LCDM$ as the mean function.
Each purple line shows a GP reconstruction with a $\chi^2$ better than the best-fit $\Lambda$CDM to the calibrated quasar dataset. The black lines correspond to the mean function.
The upper two panels show reconstructed $D_LH_0(z)$ functions and those same functions divided by the mean function ($\Lambda$CDM model that best fits the Pantheon SN Ia data). The bottom panels show $h(z)$ and om(z).}
\label{fig:QSO_GP}
\end{figure*}

\begin{figure}[t]
\centering
\includegraphics[width=0.495\textwidth]{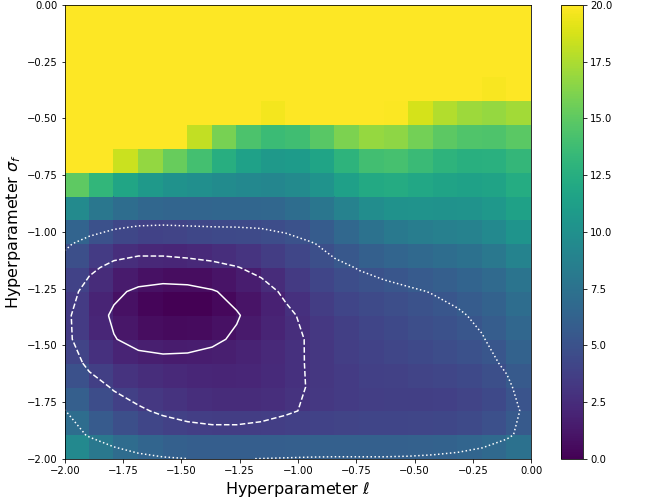}
\caption{ Posterior of the GP hyperparameters calibrated on QSO data using the the SN Ia best fitted $\LCDM$ model as a mean function. Color scale corresponds to the $-\Delta\log{\rm{Likelihood}}$, while the white contours delimit the $68.3\%$, $95.4\%$ and $99.7\%$ confidence regions.  }
\label{fig:QSO_hy}
\end{figure}

In this section we reconstruct the unanchored luminosity distances $D_LH_0$, the corresponding expansion history $h(z)$, and also the ``om diagnostic'' $om(z) = \frac{h^2(z)-1}{(1+z)^3-1}$~\citep{2008PhRvD..78j3502S} from the calibrated quasar sample with GP regression. We use the $\LCDM$ model that best fits the Pantheon data as a mean function in this GP analysis. 
These reconstruction results are shown in Figure~\ref{fig:QSO_GP}. 

From these figures we can see a noticeable evolution of these cosmological functions.  By construction, the reconstruction matches the Pantheon expectation in the redshift regime with the highest density of SN Ia (z$\lesssim 1$).  But beyond this redshift regime, the quasars prefer smaller distances than the $\Lambda$CDM fit to the Pantheon SN Ia would predict.  As can be seen in the plot of the om diagnostic (Figure~\ref{fig:best_fit_om}), the reconstructed evolution of 
this probe is arising from the same features in the data that drive shift in the inferred matter density within $\Lambda$CDM.
The evolution of the expansion history with redshift is more apparent in the GP reconstructions of Figure~\ref{fig:QSO_GP} than in the $\Lambda$CDM inference of Figure~\ref{fig:best_fit_om} basically because $\Lambda$CDM is relatively inflexible compared to GP regression. The kind of evolution in the expansion history needed to fit the full sample of the quasar data cannot be found in the $\Lambda$CDM model and shifting $\Omega_{\rm m}$ to higher values is merely a less bad fit to the data, rather than an objectively good fit.
Rather, this sort of evolution would require an evolution in the dark energy. 
We seek to be agnostic about the preferred interpretation of these reconstructions, namely, whether this indicates a beyond-$\Lambda$CDM evolution of the expansion history, some evolution in the quasar luminosity calibration hyperparameters ($\beta$, $\gamma$) or a need for additional flexibility beyond the assumed linear relation between the $\log$ of the quasars' X-ray and UV luminosities.  We leave this investigation for future work.

Rather than a visual inspection of the reconstructions, a more robust test to tell if any reconstructed, beyond-$\Lambda$CDM evolution is significant is to look at the posterior of the GP hyperparameters.
The values of the hyperparameters play an important role in GP regression, so they are not fixed to a certain value during our GP regression. In fact, the posterior of the GP hyperparameters carries important information \citep{Shafieloo2012Gaussian,2013PhRvD..87b3520S,2017JCAP...09..031A,KeeleyGW}.
If $\sigma_f$ is consistent with zero it means that there is no significant evidence for deviations from the mean function. If $\ell$ is very small this may indicate that one is over-fitting noise in the data, while if it is too large this mean that the data is uninformative about the expansion history. If the posterior for hyperparameters picks out a value for $\sigma_f$ larger than 0, then the calibrated quasar sample contains information disfavoring the mean function. In Figure~\ref{fig:QSO_hy}, we show the posterior of GP hyperparameters with flat $\LCDM$ as mean function, the results shows a mild deviation from 0, which indicates a preference for a beyond-$\LCDM$ evolution.

\section{conclusions} \label{sec:conclusion}

The quasar parameters $\gamma$, $\beta$ and $\delta$ are calibrated in a novel model-independent manner where the degeneracy between these parameters and the expansion history of the Universe is broken by using the unanchored luminosity distances $D_LH_0$ reconstructed from the Pantheon SN Ia data with GP regression and the most recent quasar sample collected in \cite{Lusso:2020pdb}. We confirm that quasars can be used as a cosmic probe based on the linear relation between $\log$ of the UV and X-ray luminosities assuming non-evolution with redshift for this relation. We also test the reliability of our calibration results by calculating the normalized residuals of $\log F_{X}$ with respect to SN Ia sample. The Gaussian distribution of the normalized residual shows the reliability of the calibration results.

Furthermore, we constrain the standard $\LCDM$ model with the calibrated quasar sample, which yields $\Omega_m = 0.382^{+0.045}_{-0.042}$ for the full sample and $\Omega_m = 0.306^{+0.046}_{-0.042}$ for the subsample. These results are consistent with previous results by \cite{Risaliti:2015zla,Risaliti:2018reu}. The expansion history can be reconstructed from the calibrated quasar sample with GP regression without any assumption about the true expansion history. The reconstructed expansion history seems evolving relative the the SN Ia predictions, especially at high redshift.  This appears to be the same evolution as seen in \cite{Lusso:2020pdb} where they find the Hubble diagram of quasars is well fit by the flat $\Lambda$CDM model at redshifts $z\sim 1.5-2$ but notice deviations occur above there.  Our model-independent (non-parametric) method that finds similar results shows additionally that this deviation is not merely a feature of the choice of parameterization but a real feature of the dataset.  

Finally, this mild preference for a beyond-$\Lambda$CDM evolution is also seen in the  posterior probability distribution of the GP hyperparameters where the preference for $\sigma_f > 0$ indicates that the data holds more information than can be modelled by the input mean function (where we assumed the best fit $\Lambda$CDM model as the mean function). Whether this information is truly indicative of beyond-$\Lambda$CDM physics or a systematic effect is uncertain, and we seek to answer this question in future works. In fact some evolution in the quasar luminosity calibration hyperparameters ($\beta$, $\gamma$) or a need for additional flexibility beyond the assumed linear relation between the $\log$ of the quasars' X-ray and UV luminosities can be also considered as alternative reasons for the observed evolution.

Summarizing, using quasars as standardizable candles can fill the redshift gap between farthest observed SN Ia and CMB measurements substantially and improve the constraints on cosmological models at $z>2$. Future surveys including \textit{Euclid}, LSST and other possible surveys will certainly provide more quasar samples. With these samples, it will be possible to have a much more precise constraint on cosmological models.

\acknowledgments

X.Li was supported by National Natural Science Foundation of China under Grants Nos. 12003006, 11947091, Hebei NSF under Grant No. A202005002 and the fund of Hebei Normal University under Grants No. L2020B20. 
A.S. would like to acknowledge the support of the Korea Institute for Advanced Study (KIAS) grant funded by the Korea government. S.Cao and Z-H.Zhu were supported by National Key R$\&$D Program of China No. 2017YFA0402600; the National Natural Science Foundation of China under Grants Nos. 12021003, 11690023, 11633001, 11920101003 and 11373014; Beijing Talents Fund of Organization Department of Beijing Municipal Committee of the CPC; the Strategic Priority Research Program of the Chinese Academy of Sciences, Grant No. XDB23000000; and the Interdiscipline Research Funds of Beijing Normal University. 
M.B. was supported by the Key Foreign Expert Program for the Central Universities No. X2018002.
This work benefits from the high performance computing clusters at College of Physics, Hebei Normal University.

\acknowledgments

\bibliography{references}

\begin{thebibliography}{}
\expandafter\ifx\csname natexlab\endcsname\relax\def\natexlab#1{#1}\fi

\bibitem[{{Aghamousa} {et~al.}(2017){Aghamousa}, {Hamann}, \&
  {Shafieloo}}]{2017JCAP...09..031A}
{Aghamousa}, A., {Hamann}, J., \& {Shafieloo}, A. 2017, \jcap, 2017, 031

\bibitem[{Aghanim {et~al.}(2020)}]{Aghanim:2018eyx}
Aghanim, N., {et~al.} 2020, Astron. Astrophys., 641, A6

\bibitem[{{Cao} {et~al.}(2017{\natexlab{a}}){Cao}, {Biesiada}, {Jackson},
  {Zheng}, {Zhao}, \& {Zhu}}]{2017JCAP0012C}
{Cao}, S., {Biesiada}, M., {Jackson}, J., {et~al.} 2017{\natexlab{a}}, \jcap,
  2017, 012

\bibitem[{{Cao} {et~al.}(2018){Cao}, {Biesiada}, {Qi}, {Pan}, {Zheng}, {Xu},
  {Ji}, \& {Zhu}}]{2018EPJC...78..749C}
{Cao}, S., {Biesiada}, M., {Qi}, J., {et~al.} 2018, European Physical Journal
  C, 78, 749

\bibitem[{{Cao} {et~al.}(2017{\natexlab{b}}){Cao}, {Zheng}, {Biesiada}, {Qi},
  {Chen}, \& {Zhu}}]{2017A&A...606A..15C}
{Cao}, S., {Zheng}, X., {Biesiada}, M., {et~al.} 2017{\natexlab{b}}, \aap, 606,
  A15

\bibitem[{Evans {et~al.}(2010)}]{Evans:2010ye}
Evans, I.~N., {et~al.} 2010, Astrophys. J. Suppl., 189, 37

\bibitem[{Foreman-Mackey {et~al.}(2013)Foreman-Mackey, Hogg, Lang, \&
  Goodman}]{foreman2013emcee}
Foreman-Mackey, D., Hogg, D.~W., Lang, D., \& Goodman, J. 2013, Publications of
  the Astronomical Society of the Pacific, 125, 306

\bibitem[{{Geng} {et~al.}(2020){Geng}, {Cao}, {Liu}, {Biesiada}, {Qi}, {Liu},
  \& {Zhu}}]{2020ApJ...905...54G}
{Geng}, S., {Cao}, S., {Liu}, T., {et~al.} 2020, \apj, 905, 54

\bibitem[{Holsclaw {et~al.}(2010{\natexlab{a}})Holsclaw, Alam, Sanso, Lee,
  Heitmann, Habib, \& Higdon}]{Holsclaw:2010sk}
Holsclaw, T., Alam, U., Sanso, B., {et~al.} 2010{\natexlab{a}}, Phys. Rev.
  Lett., 105, 241302

\bibitem[{Holsclaw {et~al.}(2010{\natexlab{b}})Holsclaw, Alam, Sanso, Lee,
  Heitmann, Habib, \& Higdon}]{Holsclaw:2010nb}
---. 2010{\natexlab{b}}, Phys. Rev. D, 82, 103502

\bibitem[{{Holsclaw} {et~al.}(2011){Holsclaw}, {Alam}, {Sans{\'o}}, {Lee},
  {Heitmann}, {Habib}, \& {Higdon}}]{2011PhRvD..84h3501H}
{Holsclaw}, T., {Alam}, U., {Sans{\'o}}, B., {et~al.} 2011, \prd, 84, 083501

\bibitem[{{Joudaki} {et~al.}(2018){Joudaki}, {Kaplinghat}, {Keeley}, \&
  {Kirkby}}]{KeeleyGP}
{Joudaki}, S., {Kaplinghat}, M., {Keeley}, R., \& {Kirkby}, D. 2018, \prd, 97,
  123501

\bibitem[{{Keeley} {et~al.}(2019){Keeley}, {Joudaki}, {Kaplinghat}, \&
  {Kirkby}}]{KeeleyTDE}
{Keeley}, R.~E., {Joudaki}, S., {Kaplinghat}, M., \& {Kirkby}, D. 2019, \jcap,
  2019, 035

\bibitem[{{Keeley} {et~al.}(2020){Keeley}, {Shafieloo}, {L'Huillier}, \&
  {Linder}}]{KeeleyGW}
{Keeley}, R.~E., {Shafieloo}, A., {L'Huillier}, B., \& {Linder}, E.~V. 2020,
  \mnras, 491, 3983

\bibitem[{Keeley {et~al.}(2021)Keeley, Shafieloo, Zhao, Vazquez, \&
  Koo}]{KeeleyeBOSS}
Keeley, R.~E., Shafieloo, A., Zhao, G.-B., Vazquez, J.~A., \& Koo, H. 2021,
  Astron. J., 161, 151

\bibitem[{Kelly(2007)}]{Kelly:2007jy}
Kelly, B.~C. 2007, Astrophys. J., 665, 1489

\bibitem[{Khadka \& Ratra(2020)}]{Khadka:2020vlh}
Khadka, N., \& Ratra, B. 2020, Mon. Not. Roy. Astron. Soc., 497, 263

\bibitem[{{Li} {et~al.}(2017){Li}, {Cao}, {Zheng}, {Qi}, {Biesiada}, \&
  {Zhu}}]{2017arXiv170808867L}
{Li}, X., {Cao}, S., {Zheng}, X., {et~al.} 2017, arXiv e-prints,
  arXiv:1708.08867

\bibitem[{{Liao} {et~al.}(2019){Liao}, {Shafieloo}, {Keeley}, \&
  {Linder}}]{KeeleySLa}
{Liao}, K., {Shafieloo}, A., {Keeley}, R.~E., \& {Linder}, E.~V. 2019, \apjl,
  886, L23

\bibitem[{{Liao} {et~al.}(2020){Liao}, {Shafieloo}, {Keeley}, \&
  {Linder}}]{KeeleySLb}
---. 2020, \apjl, 895, L29

\bibitem[{Liu {et~al.}(2020)Liu, Cao, Biesiada, Liu, Geng, \&
  Lian}]{Liutonghua_20201}
Liu, T., Cao, S., Biesiada, M., {et~al.} 2020, The Astrophysical Journal, 899,
  71

\bibitem[{{Liu} {et~al.}(2020{\natexlab{a}}){Liu}, {Cao}, {Zhang}, {Biesiada},
  {Liu}, \& {Lian}}]{2020MNRAS.496..708L}
{Liu}, T., {Cao}, S., {Zhang}, J., {et~al.} 2020{\natexlab{a}}, \mnras, 496,
  708

\bibitem[{{Liu} {et~al.}(2020{\natexlab{b}}){Liu}, {Cao}, {Liu}, {Li}, {Geng},
  {Lian}, \& {Guo}}]{Liu:2020pfa}
{Liu}, Y., {Cao}, S., {Liu}, T., {et~al.} 2020{\natexlab{b}}, \apj, 901, 129

\bibitem[{Lusso(2020)}]{Lusso:2020obu}
Lusso, E. 2020, Front. Astron. Space Sci., 7, 8

\bibitem[{Lusso {et~al.}(2019)Lusso, Piedipalumbo, Risaliti, Paolillo, Bisogni,
  Nardini, \& Amati}]{Lusso:2019akb}
Lusso, E., Piedipalumbo, E., Risaliti, G., {et~al.} 2019, Astron. Astrophys.,
  628, L4

\bibitem[{Lusso \& Risaliti(2016)}]{lusso2016tight}
Lusso, E., \& Risaliti, G. 2016, The Astrophysical Journal, 819, 154

\bibitem[{Lusso \& Risaliti(2017)}]{Lusso:2017hgz}
---. 2017, Astron. Astrophys., 602, A79

\bibitem[{Lusso {et~al.}(2020)}]{Lusso:2020pdb}
Lusso, E., {et~al.} 2020, Astron. Astrophys., 642, A150

\bibitem[{Menzel {et~al.}(2016)Menzel, Merloni, Georgakakis, Salvato, Aubourg,
  Brandt, Brusa, Buchner, Dwelly, Nandra, Pâris, Petitjean, \&
  Schwope}]{10.1093/mnras/stv2749}
Menzel, M.-L., Merloni, A., Georgakakis, A., {et~al.} 2016, Monthly Notices of
  the Royal Astronomical Society, 457, 110

\bibitem[{Mingo {et~al.}(2016)Mingo, Watson, Rosen, Hardcastle, Ruiz, Blain,
  Carrera, Mateos, Pineau, \& Stewart}]{Mingo:2016pdq}
Mingo, B., Watson, M., Rosen, S., {et~al.} 2016, Mon. Not. Roy. Astron. Soc.,
  462, 2631

\bibitem[{Mortlock {et~al.}(2011)}]{Mortlock:2011va}
Mortlock, D.~J., {et~al.} 2011, Nature, 474, 616

\bibitem[{Nardini {et~al.}(2019)Nardini, Lusso, Risaliti,
  {et~al.}}]{Nardini:2019}
Nardini, E., Lusso, E., Risaliti, G., {et~al.} 2019, Astron. Astrophys., 632,
  A109

\bibitem[{Pâris {et~al.}(2017)}]{Paris:2016xdm}
Pâris, I., {et~al.} 2017, Astron. Astrophys., 597, A79

\bibitem[{{Qi} {et~al.}(2019){Qi}, {Cao}, {Zhang}, {Biesiada}, {Wu}, \&
  {Zhu}}]{2019MNRAS.483.1104Q}
{Qi}, J.-Z., {Cao}, S., {Zhang}, S., {et~al.} 2019, \mnras, 483, 1104

\bibitem[{Qi {et~al.}(2021)Qi, Zhao, Cao, Biesiada, \& Liu}]{Qi_2021}
Qi, J.-Z., Zhao, J.-W., Cao, S., Biesiada, M., \& Liu, Y. 2021, Monthly Notices
  of the Royal Astronomical Society, doi:10.1093/mnras/stab638

\bibitem[{Rasmussen \& Williams(2006)}]{Rasmussen:2006}
Rasmussen, C.~E., \& Williams, C. K.~I. 2006, The MIT Press,
  http://www.gaussianprocess.org/gpml/

\bibitem[{Risaliti \& Lusso(2015)}]{Risaliti:2015zla}
Risaliti, G., \& Lusso, E. 2015, Astrophys. J., 815, 33

\bibitem[{Risaliti \& Lusso(2017)}]{Risaliti:2016nqt}
---. 2017, Astron. Nachr., 338, 329

\bibitem[{Risaliti \& Lusso(2019)}]{Risaliti:2018reu}
---. 2019, Nature Astron., 3, 272

\bibitem[{{Sahni} {et~al.}(2008){Sahni}, {Shafieloo}, \&
  {Starobinsky}}]{2008PhRvD..78j3502S}
{Sahni}, V., {Shafieloo}, A., \& {Starobinsky}, A.~A. 2008, \prd, 78, 103502

\bibitem[{Salvestrini {et~al.}(2019)Salvestrini, Risaliti, Bisogni, Lusso, \&
  Vignali}]{Salvestrini:2019thn}
Salvestrini, F., Risaliti, G., Bisogni, S., Lusso, E., \& Vignali, C. 2019,
  Astron. Astrophys., 631, A120

\bibitem[{Scolnic {et~al.}(2017)Scolnic, Jones, Rest, Pan, Chornock, Foley,
  Huber, Kessler, Narayan, Riess, {et~al.}}]{scolnic2017complete}
Scolnic, D., Jones, D., Rest, A., {et~al.} 2017, arXiv preprint
  arXiv:1710.00845

\bibitem[{Shafieloo {et~al.}(2012)Shafieloo, Kim, \&
  Linder}]{Shafieloo2012Gaussian}
Shafieloo, A., Kim, A.~G., \& Linder, E.~V. 2012, Physical Review D, 85, 123530

\bibitem[{{Shafieloo} {et~al.}(2013){Shafieloo}, {Kim}, \&
  {Linder}}]{2013PhRvD..87b3520S}
{Shafieloo}, A., {Kim}, A.~G., \& {Linder}, E.~V. 2013, \prd, 87, 023520

\bibitem[{Webb {et~al.}(2020)}]{Webb:2020rgy}
Webb, N.~A., {et~al.} 2020, Astron. Astrophys., 641, A136

\bibitem[{{Zheng} {et~al.}(2021){Zheng}, {Cao}, {Biesiada}, {Li}, {Liu}, \&
  {Liu}}]{Zheng:2021}
{Zheng}, X., {Cao}, S., {Biesiada}, M., {et~al.} 2021, arXiv e-prints,
  arXiv:2103.07139

\end{thebibliography}
\end{document}